\documentclass[12pt,a4paper,leqno]{article}
\title{Integrability of differential equations with fluid mechanics application: from 
Painleve property to the method of simplest equation}
\author{Zlatinka I. Dimitrova$^1$, Kaloyan N. Vitanov$^2$}
\date{$^1$ 'G. Nadjakov' Institute of Solid State Physics, Bulgarian Academy
of Sciences, Blvd. Tzarigradsko Chaussee 71, 1784 Sofia, Bulgaria \\
$^2$ Faculty of Mathematics and Informatics, 'St. Kliment Ohridski' University of 
Sofia, Blvd. J. Bourchier 5, 1164 Sofia, Bulgaria}
\begin{document}
\maketitle
\begin{abstract}
We present a brief overview of integrability of nonlinear ordinary and partial
differential equations with a focus on the Painleve
property: an ODE of second order has the Painleve property if the only
movable singularities connected to this equation are single poles. The importance
of this property can be seen from the Ablowitz-Ramani-Segur  conhecture that states
that a nonlinear PDE is solvable by inverse scattering transformation only if
each nonlinear ODE obtained by exact reduction of this PDE has the Painleve
property. The  Painleve property motivated motivated much research on obtaining 
exact solutions on nonlinear PDEs and leaded in particular to the method of simplest equation.
A version of this method called modified method of simplest equation is discussed below.
\end{abstract}
\newpage
\section{Several words on the integrability of differential equations}
The development of the modern natural sciences leaded to wide use of
nonlinear models for  natural and social phenomena \cite{q1}-\cite{q4}.
Thus the problems connected to obtaining exact solutions of the model 
nonlinear differential equations are very actual \cite{p1}-\cite{p5}.
For many researchers the question about the integrability of nonlinear differential equations
or systems of such equations is not very clear. They have probably heard is
that there exist a complicated method called Inverse Scattering Transformation
(IST) \cite{a1,a2} that allows obtaining exact solutions of some nonlinear PDEs. May be the
researchers have heard that integration by IST can be
performed if the corresponding nonlinear differential equation is
somehow connected to a property called Painleve property that states 
something about the singularities of the solution in the complex plane.
Many researchers are puzzled by this fact as they have to solve differential
equations which contain real variables.  We shall discuss below the connection between the 
methodology of the test for the
Painleve property and the research on direct methods for
obtaining exact solutions of nonlinear differential equations.
\par
Given an differential equation or a system of equations how one can tell 
appriori whether or not they are integrable?
For a second order ordinary differential equation the existence of a constant
first integral (which for an example is the energy for the many problems connected
to the physics) leads to a reduction of the order of the equation by $1$. Thus the
task for finding a solution is reduced to quadrature.
For second order systems especially if they not conservative the integration is 
more difficult. Such  systems possess solution but the problem for finding analytical
representation of the solution is very difficult to solve. It is
extremely useful to find integrals of motion connected to the solved equations.
But how to find integrals of motion if they exist? The existence of first integrals may be
connected to the analytical structure of the studied differential equation as it was implied
by the research of Sofia Kovalevskaya. 
\section{The research of Kovalevskaya and Painleve}
Kovalevskaya won the Bordin prize of the Paris Academy of Sciences in 1888 for the solution of
the system of six first-order nonlinear ODEs that describe the motion of a heavy top rotation
about a fixed point. Kovalevskaya solved this mechanical problem by the methods
of complex variable theory (she  obtained
conditions under which the only movable singularities of the solutions in the complex plane are
single poles). Movable singularity is a singularity whose
location depends on the initial conditions. Let us consider the simple nonlinear equation in the
complex plane \cite{tabor} ($z$ below denotes a complex number)
\begin{equation}\label{b1}
\frac{df}{dz} + f^2(z) =0.
\end{equation}
It has the solution
\begin{equation}\label{b2}
f(z) = \frac{1}{z+z_0}; \ \ z_0 = \frac{1}{f(0)}.
\end{equation}
In other words $f(z)$ has a singularity at $z=z_0$ and this singularity is movable as it depends on the
initial condition. Painleve continued this research on the connection between analytic structure
and integrability by analysis of the class of  second order differential equations
\begin{equation}\label{b3}
\frac{d^2 y}{dz^2} = F \left(\frac{dy}{dz}, y, z \right),
\end{equation}
where $F$ is analytic function in $z$ and rational in $y$ and $\frac{dy}{dz}$. Painleve obtained
$50$ types of such equations whose only movable singularities were simple poles. $44$ of these equations
can be integrated in terms of known functions. The remained six equations are called Painleve 
transcendents and they can not be integrated by quadratures. The simplest two Painleve transcendents are
\begin{flushleft}
{\rm P1:} \ \  $\frac{d^2y}{dx^2}=6y^2 +x$; \ 
{\rm P2:} \ \ $\frac{d^2y}{dx^2}=2y^3+xy + \pi$; \ \ $\pi$: \ {\rm free parameter} .
\end{flushleft}
\par
The solutions even for simple ODEs in complex plane are not known but the nature of their movable
singularities can be determined by studying the local properties of the solutions. Let as above
the function $F$ is analytic in $z$ and rational in other arguments. The behavior of the movable
singularities of the $n$-th order ODE
\begin{equation}\label{b4}
\frac{d^n y}{dz^n} = F \left( \frac{d^{n-1}y}{dz^{n-1}},\dots,\frac{dy}{dz},y,z \right)
\end{equation}
can be determined by leading order analysis, i.e., by the ansatz
\begin{equation}\label{b5}
y(z) = a(z-z_0)^\alpha.
\end{equation}
Let us discuss an equation which contains as particular case the first Painleve transcendent
($A$ below is a parameter) \cite{tabor}
\begin{equation}\label{b6}
\frac{d^2y}{dz^2}=6y^2 + Az.
\end{equation}
We substitute Eq.(\ref{b5}) in Eq.(\ref{b6}) and obtain as a result a relationship that
contains the powers $\alpha -2$; $2 \alpha$; and $\alpha$. In order to obtain a non-zero expression
of kind (\ref{b5}) we have to balance at least two of the powers. Taking into an account that
the highest order singularity arises from the lowest (negative) value of $\alpha$ the
appropriate balance is:
$\alpha - 2 = 2 \alpha \ \to \alpha = -2$.
Then at $z$ close to $z_0$ the leading order of the solution is $\frac{1}{(z-z_0)^2}$ i.e. second
order (movable simple) pole. In order to characterize better the behavior in the neighborhood
of the singularity a local expansion of the solution has to be considered. If the singularity
is a movable pole the expansion is a Laurent series. In our case (the second-order pole above)
the candidate for Laurent series is
\begin{equation}\label{b8}
y(z) = \sum_{i=0}^{\infty} a_i \left(z - z_0 \right)^{i-2}.
\end{equation}
After the substitution of Eq.(\ref{b8}) in the equation (\ref{b6}) one obtains a system of
recursion equations ($a_0 =1$):
$a_i (i+1)(i-6) = 6 \sum_{l=1}^{i-1} a_{i-l} a_l + A a_{i-2}$.
The solution of the system of equations is: $a_1 =0;\  a_2 = -\frac{A}{12};\  a_3 =0; 
a_4 = - \frac{A^2}{24}$; 
$a_5 =0 ; \ 0.a_6 = 12 a_2 a_4 + A a_4 =0$.
Thus the series (\ref{b8}) have two free parameters $a_6$ 
and $z_0$ (the arbitrariness of the pole position). As the investigated ODE (\ref{b6})
is of second order its general solution contains two free parameters. These free parameters
are contained in the local series expansion, i.e., (\ref{b8}) are Laurent series of the general
solution indeed. In addition in the neighborhood
of the movable singularity $z_0$ the solution of Eq.(\ref{b6}) behaves as second order
pole, i.e., the equation has the Painleve property (which is a strong hint for its
integrability).
\par
The powers of $(z-z_0)$ at which the arbitrary coefficient appear are called resonances.
At the resonances one obtains a compatibility condition which must be satisfied in order to
ensure the arbitrariness of the corresponding coefficient. In our case the resonance occurred
at $i=6$ and the compatibility condition is: $ 12 a_2 a_4 + A a_4 =0$.
\par
The leading order of the solution behavior near the pole as well as the value of the
resonance can be integer number (and then the pole is simple and the equation has
the Painleve property) but these two numbers can be also non-integer (for an example
irrational or even complex). In this case the singularity is not a simple pole
and the equation does not have the Painleve property (which is a hint of non-integrability
from the point of view of the research of Kovalevskaya and Painleve).
\par
In some cases the Painleve property can be established only for selected values of
the parameters of the studied nonlinear ODEs. In such case one can attempt the $\Psi$-series
technique that will not be discussed here.
\section{Weiss-Tabor-Carnevale research on Painleve property for nonlinear PDEs}
Weiss, Tabor, and Carnevale \cite{wtk} continued the research of Painleve in the direction of
finding hints for obtaining exact solutions of nonlinear partial differential equations.
They work was motivated by the conjecture of Ablowitz, Ramani and Segur \cite{abl1}
that a nonlinear PDE is solvable by inverse scattering transform (IST) only if each nonlinear
ODE obtained by exact reduction of this PDE has the Painleve property. In other words we have to
test all reductions of the PDE to ODEs for Painleve property and if all ODEs have the
Painleve property then the corresponding nonlinear PDE has to be integrable by IST. But what are
all these reductions? We don't know them. Then one has to follow another way and one possible way
the the generalized Laurent expansion approach of Weiss, Tabor and Carnevale. The key point in
this approach is that as difference to the case of complex function of one complex variable for
the case of functions of several complex variables the singularities can't be isolated and these
singularities are located in singular manifolds. If $f(z_1,\dots,z_n)$ is a meromorphic function
of the complex variables $z_1,\dots,z_n$ the singularities of $f$ are located in analytic
manifolds (called singular manifolds) of dimension $2n-2$ which are determined by conditions like
\begin{equation}\label{c1}
\psi(z_1,\dots,z_n)=0 
\end{equation}
where $\psi$ is analytic function in the neighborhood of the manifold
defined by Eq.(\ref{c1}). Weiss, Tabor and Carnevale have proposed generalized Laurent
expansion of analytic function $f(z_1,\dots,z_n)$ as follows
\begin{equation}\label{c2}
f(z_1,\dots,z_n) = \frac{1}{\phi^\alpha} \sum_{i=0}^\infty u_i \phi^i
\end{equation}
where $\phi(z_1,\dots,z_n)$ and $u_i(z_1,\dots,z_n)$ are analytic functions of $z_1, \dots,z_n$
in the neighborhood of the manifold given by Eq.(\ref{c1}). $\alpha$ is an integer. 
Difference to the classic Laurent series is that instead of $z-z_0$ we have the function 
$\phi$ and $u_i$ are functions and not constant coefficients as in the classic case. Additional 
requirement to the function $\phi$ is that its gradients do not vanish on the singular manifold.
\par
The test for the generalized Painleve property happens as in the case of ordinary differential equations.
First Eq.(\ref{c2}) is substituted in the studied nonlinear PDE and the possible values of $\alpha$ and 
the recursion relations for the function $u_i$ are determined. The recursion relations are
coupled PDFs containing the functions $\phi$ and $u_i$. As in the case of ODEs discussed above
there will be powers of $\phi$ for which the corresponding $u_i$ are arbitrary (resonances).
If some $u_i$ are not arbitrary at the resonance some kind of generalized $\Psi$-series can be introduced
to ensure their arbitrariness. 
\section{Application to systems connected to fluid mechanics}
We shall discuss the application the methodology for detecting Painleve property to 
the Lorenz system and one of its generalizations, namely
\begin{eqnarray}\label{f1}
\frac{dx}{dt} &=& - \sigma x + \sigma y - \omega yz, \nonumber\\
\frac{dy}{dt} &=& rx - y - m xz, \  \frac{dz}{dt} = -bz + xy.
\end{eqnarray}
where $\sigma$, $\omega$, $r$, $m$, and $b$ are parameters. When $\omega = 0$;  $m = 1$ the system
(\ref{f1}) is reduced to the classical Lorenz system well known from meteorology, nonlinear
dynamics and chaos theory. The classic Lorenz system was tested for presence of 
Painleve property in \cite{t1}. 
Let us transform the equations of the system (\ref{f1}) by setting
$
x=\frac{X}{\epsilon}; \ y = \frac{Y}{\sigma \epsilon^2}; \ z = \frac{Z}{\sigma \epsilon^2};
\ t= \epsilon T; \ \epsilon = \frac{1}{\sqrt{\sigma r}}.
$
The system (\ref{f1}) becomes
\begin{eqnarray}\label{f2}
\frac{dX}{dT} &=& - \sigma \epsilon X +  Y - \frac{\omega}{\sigma^2 \epsilon} YZ, \nonumber\\
\frac{dY}{dT} &=& X - \epsilon Y - mXZ, \ 
\frac{dZ}{dT} = -b \epsilon Z + XY.
\end{eqnarray}
We shall investigate the system of equations (\ref{f2}) for presence of Painleve property.
\par
First of all we have to perform the leading order analysis by setting
\begin{equation}\label{f3}
X= \frac{a}{T^\alpha}; \ Y = \frac{b}{T^\beta}; \ Z = \frac{c}{T^\gamma},
\end{equation}
where $a, b, c$ and $\alpha, \beta, \gamma$ are parameters. The substitution of
Eqs. (\ref{f3}) in the system (\ref{f2}) leads to the following relationships
after balancing the largest powers in the denominators of the resulting three
equations: $\alpha = \beta = \gamma = 1$.
This result is different from the case of classical Lorenz system where $m=1$ and $\omega = 0$
(the last relationships leads to significant changes of the right-hand side of the
first equation from (\ref{f2}) - the nonlinearity there is removed). For this case
the substitution of Eqs. (\ref{f3}) in the system (\ref{f2}) leads to the result
$\alpha = 1; \ \beta = \gamma = 2$ .
For the case of generalized Lorenz system we obtain the following system
of equation for the parameters $a, b, c$ (after equating the numerators of
the two largest powers from the denominators from the equations for the
leading order analysis): 
\begin{eqnarray}\label{f6}
- a + \frac{\omega b c}{\sigma^2 \epsilon} = 0,  \
- b + m a c = 0, \
- c - a b = 0.
\end{eqnarray}
The non-trivial solution of the system of equations (\ref{f6}) is
$
a = \pm \frac{i}{\sqrt{m}}; \ b= \pm i \sqrt{\frac{\epsilon}{\omega}}; \ c= \sqrt{\frac{\epsilon}{m \omega}}
$.
\par
What follows are the power series expansions around the singularities.
On the basis of the above results we shall consider the following expansions
for the three unknown functions around the singularity of the solution which
we assume to be at $(X_0, Y_0, Z_0)$:
\begin{eqnarray}\label{f7}
X &=& \frac{i}{\sqrt{m} (T-T_0)} \sum_{j=0}^\infty a_j (T-T_0)^j, \
Y = \frac{i \sqrt{\epsilon}}{\sqrt{\omega} (T-T_0)} \sum_{j=0}^\infty b_j (T-T_0)^j,  \nonumber \\
Z &=& \frac{ \sqrt{\epsilon}}{\sqrt{ m \omega} (T-T_0)} \sum_{j=0}^\infty c_j (T-T_0)^j.
\nonumber \\
\end{eqnarray}
These series are different from the series for the classical Lorenz system. For the
classical Lorenz system the system of equations for the parameters $a, b, c$ is
\begin{eqnarray}\label{f8}
a +  b = 0, \
a c - 2 b = 0, \
2 c + a b = 0.
\end{eqnarray} 
which solution is
$
a = \pm 2 i; \ b = \mp 2 i; \ c = -2,
$
and the expansions for $X, Y, Z$ around the singularity become
\begin{eqnarray}\label{f10}
X &=& \frac{2 i}{(T-T_0)} \sum_{j=0}^\infty a_j (T-T_0)^j, \
Y = \frac{- 2 i}{ (T-T_0)^2} \sum_{j=0}^\infty b_j (T-T_0)^j, \nonumber \\
Z &=& \frac{ - 2}{ (T-T_0)^2} \sum_{j=0}^\infty c_j (T-T_0)^j.
\end{eqnarray}
\par
The next step is obtaining of the recurrence equations.
We have to substitute the expansion (\ref{f7}) in the generalized Lorenz
system (\ref{f1}) and by equalizing to $0$ the coefficients of the powers $(T-T_0)^{-n}$,
$n=-1,0,1,\dots$ we shall obtain a system of nonlinear recurrent equations for the
coefficients $a_j$, $b_j$ and $c_j$. Setting to $0$ the coefficients of $(T-T_0)^{-2}$ in the
resulting equations we obtain that
$
a_0 = b_0 = c_0 = 1.
$
 Setting to $0$ the coefficients of $(T-T_0)^{-1}$ in the
resulting equations we obtain the following system for the coefficients $a_1$, $b_1$, $c_1$:
\begin{eqnarray}\label{f12}
\sigma \sqrt{\epsilon m} (\sqrt{\epsilon} - \sqrt{\omega}) + c_1 + b_1 = 0, \
\sigma \epsilon + c_1 + a_1 \sigma - \sqrt{\omega}{m \epsilon} = 0, 
\nonumber \\
a_1 + b_1 + \epsilon b = 0.
\end{eqnarray}
The solution of this system is
\begin{eqnarray}\label{f13}
a_1 &=& \frac{1}{2} \left[-\epsilon + \sigma \sqrt{\frac{\omega}{\epsilon m}} - \sigma \sqrt{\frac{m}{\epsilon \omega}} - \sigma \epsilon - \epsilon b  \right],
\nonumber \\
b_1 &=& - \bigg[ \sigma \epsilon - \sigma \sqrt{\frac{\epsilon m}{\omega}} - \frac{1}{2} \bigg(\epsilon - 
 \frac{1}{\sigma} \sqrt{\omega}{\epsilon m} - \sigma \sqrt{\frac{m \epsilon}{\omega}} + \sigma \epsilon - b \epsilon \bigg) \bigg],
\nonumber \\
c_1 &=& \frac{1}{2} \bigg[-\epsilon + \sigma \sqrt{\frac{\omega}{\epsilon m}} + \sigma \sqrt{\frac{m \epsilon}{\omega}} - 
 \sigma \epsilon + \epsilon b \bigg].
\end{eqnarray}
\par
The situation with classic Lorenz system ($\omega =0$, $m=1$) is different again.
In this case we have to substitute the series (\ref{f10}) in (\ref{f1}) (with
$\omega = $ and $m=1$). Setting to $0$ the coefficient of $(T-T_0)^{-2}$ in the
first resulting equation and the coefficients of $(T-T_0)^{-3}$ in the second ant third
resulting equation we obtain the following result
$a_0 = b_0 = c_0 =1$.
Next we have to set to $0$ the coefficient of $(T-T_0)^{-1}$ in the
first resulting equation and the coefficients of $(T-T_0)^{-1}$ in the second ant third
resulting equation. We obtain the following system of equations for $a_1$, $b_1$ and
$c_1$:
$$
2 \sigma \epsilon i + 2  i b_1 = 0, \
b_1 -  \epsilon  - 2  c_1 - 2  a_1 = 0, \
c_1 - 2  b_1 - 2  a_1 -  \epsilon b = 0.
$$
The solution of this system is 
\begin{eqnarray}\label{l3}
a_1 = \frac{(3 \sigma - 2 b - 1)\epsilon }{6}, \
b_1 = - \epsilon \sigma, \
c_1 = \frac{b - 1 - 3 \sigma}{3}.
\end{eqnarray}
For $j=2,3,\dots$ for the generalized Lorenz system we obtain the system of recurrence equations
\begin{eqnarray}\label{l3a}
\frac{1}{\sqrt{m}}(j-1)a_j+ \frac{\sigma \epsilon}{\sqrt{m}}a_{j-1}-
\frac{\sqrt{\epsilon}}{\sqrt{\omega}}b_{j-1} + 
\frac{1}{m \sigma^2} \sum_{l=0}^j b_l c_{j-l} =0, \nonumber \\
\frac{\sqrt{\epsilon}}{\sqrt{\omega}}(j-1) b_j - \frac{1}{\sqrt{m}}a_{j-1} + 
\frac{\epsilon \sqrt{\epsilon}}{\sqrt{\omega}} b_{j-1} +
\frac{\sqrt{\epsilon}}{\sqrt{m} \sqrt{\omega}} \sum_{l=0}^j a_l c_{j-l}=0, \nonumber \\
\frac{1}{m}(j-1)c_j + \frac{b \epsilon}{m} c_{j-1} + \frac{1}{\sqrt{m}} \sum_{l=0}^j a_l b_{j-l}=0.
\end{eqnarray}
\par
Let us obtain the first relationship from the system (\ref{l3a}) in more detail.
The substitution of the series (\ref{f7}) in the system of generalized Lorenz
equations (\ref{f2}) leads to the relationship
\begin{eqnarray}\label{l3b}
\sum_{j=0}^\infty (T-T_0)^{j-2} \bigg[\frac{1}{\sqrt{m}}(j-1) a_j + \frac{\sigma \epsilon}{\sqrt{m}} a_{j-1} - 
\nonumber \\
\frac{\sqrt{\epsilon}}{\sqrt{\omega}} b_{j-1} \bigg] +  
\frac{1}{m \sigma^2} \sum_{l=0}^\infty \sum_{k=0}^\infty b_l c_k (T-T_0)^{l+k-2} =0. 
\end{eqnarray}
From Eq.(\ref{l3b}) we want to select the terms of the same power of $(T-T_0)$.
This means that we must have $j-2 = l+k-2$, i.e., $j=l+k$. In other words from
the double sum in (\ref{l3b}) we have to select the terms for which $k=j-l$ and
in addition we have to take into an account that $k \ge 0$ which leads to cutting the sum
with respect to $l$ from $l=0$ to $l=j$. Then the terms from the double sum
that participate in the recurrence relationship are given by ($\delta_{i,j}$
is the delta-symbol of Kronecker)
\begin{eqnarray}\label{l3c}
\sum_{l=0}^j \sum_{k=0}^\infty b_l c_k \delta_{k,j-l} (T-T_0)^{l+ k -2} \to
\sum_{l=0}^j b_l c_{j-l} (T-T_0)^{j-2}.
\end{eqnarray}
We note that in the last relationship $j$ was fixed. Putting all together for
fixed value of $j$ in Eq.(\ref{l3b}) we obtain the first relationship from (\ref{l3a}).
The remaining relationships from (\ref{l3a}) are obtained in the same way.
\par
For comparison the situation in the classic Lorenz system is simpler.
For $j=2,3,\dots$ we obtain the system of recurrence equations
\begin{eqnarray}\label{l3}
(j-1)a_j + b_j + \sigma \epsilon a_{j-1} =0, \nonumber \\
2 a_j + (j-2) b_j + 2 c_j + 2 \sum_{k=1}^{j-1} a_{j-k} c_k + 
a_{j-2} +
\epsilon b_{j-1}=0, \nonumber \\
2 a_j + 2 b_j + (j-20 c_j + 2 \sum_{k=1}^{j-1} a_{j-k} b_k + 
\epsilon b c_{j-1} =0.  
\end{eqnarray}
For $j=2$ and $j=4$ there are resonances that lead to the following
relationships among the parameters of the equations of the classic Lorenz system
\begin{eqnarray}\label{l4}
6 \sigma^2 - \sigma b - 2 \sigma - b(b-1) = 0, \nonumber \\
(b-1)[24 + 57(\sigma -1) - 15(b-1)] -
9 \sigma (2 \sigma -1) =0, \nonumber \\
6(1+ b - \sigma) a_1^2 c  + \epsilon (2 \sigma - b - 5)\bigg ( 
2 \sigma a_1^2 +
6 a_1 c_1\frac{b(1-\sigma \epsilon^2)}{2}\bigg) =0.
\end{eqnarray}
The last relationships fix conditions on the values of $\epsilon$, $\sigma$ 
and $\beta$ for which one the classical Lorenz system is integrable.
As $\epsilon$ participates only in the third equations from (\ref{l4})
from the first two relationships there one determined $\sigma$ and $b$
and then the last relationship leads to the corresponding value of
$\epsilon$. One example for such solution is $b=1$; $\sigma = 1/2$,
$\epsilon = \infty$. Then the solution of the Lorenz system can be
constructed on the basis of elliptic functions. We let the interested reader to
find the values of parameters for which the generalized Lorenz system is
integrable.
\section{Method of simplest equation and its version called modified method
of simplest equation}
From the many approaches for obtaining exact analytical solutions 
of nonlinear PDEs here we shall give short remarks on the method of
simplest equation \cite{k1, k2} and on his version called
modified method of simplest equation \cite{m1}-\cite{m3}.
The method of simplest equation uses the first step of the test for the Painleve
property: the leading order analysis to determine the truncation of the
sum from Eq.(\ref{et3}) below. In the modified method of simplest equation
this first step is changed by the equivalent procedure of solving of balance
equation. The methodology
is based on the fact that after application
of appropriate ansatz large class of  NPDEs can be reduced to ODEs of the kind
\begin{equation}\label{et2}
{\cal{P}}\left( F(\xi),\frac{d F}{d \xi}, \frac{d^2F}{d \xi^2},\dots \right)=0,
\end{equation}
and for some equations of the kind (\ref{et2}) particular solutions can be obtained
which are finite series
\begin{equation}\label{et3}
F(\xi) = \sum_{i=1}^P a_i [\Phi(\xi)]^i,
\end{equation}
constructed by solution $\Phi(\xi)$ of more simple equation referred to as
simplest equation. The simplest equation can be the equation of Bernoulli, equation
of Riccatti, etc. The substitution of Eq. (\ref{et3}) in Eq. (\ref{et2})
leads to the polynomial equation
\begin{equation}\label{et4}
{\cal P}= \sigma_0 + \sigma_1 \Phi + \sigma_2 \Phi^2 + \dots + \sigma_r \Phi^r =0 ,
\end{equation}
where the coefficients $\sigma_i$, $i=0,1,\dots,r$ depend on the parameters of the
equation and on the parameters of the solutions. Equating all these coefficients to $0$,
i.e.,  by setting
\begin{equation}\label{et5}
\sigma_i =0, i=1,2,\dots,r ,
\end{equation}
one obtains a system of nonlinear algebraic equations. Each solution of this system
leads to a solution of kind (\ref{et3}) of Eq. (\ref{et2}).
\par
In order to ensure non-trivial solution by the above method we have to ensure 
that $\sigma_r$ contains at least two terms. To do this in the modified simplest
equation method we have to balance the highest
powers of $\phi$ that are obtained from the different terms of the solved equation of
kind (\ref{et2}). As a result of this we obtain an additional equation between some of the
parameters of the equation and the solution. This equation is called balance 
equation \cite{m1}-\cite{m3}.
\par
As conclusion we stress that there is a unproved conjecture that the Painleve 
property is a direct test of integrability
of a differential equation or for a system of differential equations.  
There is however a solid evidence that the conjecture is true \cite{tabor}. The
Liouville theorem from the complex analysis states that if $f(z)$ is entire bounded 
function of the complex variable $z$ the it can take only one (constant) value
$f(z) = c$.
Example of such entire bounded functions (if we think the time $t$ to be complex variable)
are the time-independent integrals of the motion connected to a differential equation or
to a system of such equations. Let us assume that the studied system of equations is
Hamiltonian and algebraic integrable, i.e., the time-independent integrals are polynomials
of the canonical variables $p_i$ and $q_i$. In the complex plane the functions $p_i$ and
$q_i$ can have various kinds of movable singularities, but...., as the constructed by them
integral has to be a constant in the complex plane all the singularities must cancel at any
singularity position $t_0$. But this is possible for polynomial in $p_i$ and $q_i$ integrals 
only if the singularities of $p_i$ and $q_i$ are simple poles or rational branch points. 
For other kinds of terms in the integrals it will be not possible to construct a polynomial 
function in which the corresponding power of singularities cancel. The words above hint that
if the studied system has Hamiltonian with integrals that are not polynomials (and are
for an example irrational or transcendental functions) the Painleve property could not be
considered as appropriate test for integrability of corresponding system of differential equations.

\end{document}